\documentclass[12pt,a4paper]{article}
\usepackage[british]{babel}
\usepackage{epsfig}
\usepackage{amsmath}
\usepackage{cite}
\usepackage{subfigure}
\usepackage{algorithm}
\usepackage{algorithmicx}
\usepackage{algpseudocode}
\usepackage{hyperref}
\title{
  {\vspace*{-6mm} \normalsize
    \hfill \parbox{40mm}{DESY 11-076     \\[25mm] }}\\[-12mm]
  Exploratory investigation of nucleon-nucleon interactions\\
  using Euclidean Monte Carlo simulations}
\author{ Istvan Montvay$^{a,b}$ and Carsten Urbach$^{b}$\\[8mm]
 {\small $^a$ Deutsches Elektronen-Synchrotron DESY, Notkestr.\,85,
  D-22603 Hamburg, Germany}\\
 {\small $^b$ HISKP (Theorie), Universit\"at Bonn, Nussallee 14-16,
  D-53115 Bonn, Germany}\\[8mm] }      
\date{May, 2011}

\begin{document}
\maketitle

\begin{abstract}
  We present an exploratory study of chiral effective theories of
  nuclei with methods adopted from lattice quantum chromodynamics
  (QCD).
  We show that the simulations in the Euclidean path integral
  approach are feasible and that we can determine the energy of the
  two nucleon state.
  This opens up the possibility to determine in future simulations
  nucleon phase shifts by varying the parameters and the simulated
  volumes.
  The physical cut-off of the theory is realised by blocking
  of the lattice fields.
  By keeping the block size fixed in physical units the lattice
  cut-off (i.e. the lattice spacing) can be freely changed.
  This offers an effective way for controlling lattice artefacts. 
\end{abstract}
\newpage

\section{Introduction}

Quantum Chromodynamics (QCD) is the nowadays accepted theory of strong
interactions in terms of the fundamental quark and gluon degrees of
freedom. Also, if one is interested in nuclear physics,
QCD is the relevant theory to calculate for instance binding energies
of nuclei from first principles.  However, QCD is strongly interacting
at low energies and hence, non-perturbative methods are needed in
order to study quantities as, for instance, nuclear binding energies.

The method of choice for ab initio, non-perturbative investigations of
QCD is provided by lattice QCD. Even if this field was facing
tremendous progress over the last few years, the computation of
nuclear binding energies for say $A>2$ is still out of reach. But it
is also not clear whether those quantities need to be computed in the
fundamental theory, because the relevant degrees of freedom are
effectively given by hadrons described in a chiral effective
theory~\cite{Weinberg:1990rz,Weinberg:1991um}, 
and not by quarks and gluons in QCD. And again, lattice methods can be
used in order to investigate these chiral effective theories from
first principles, for a review see ref.~\cite{Lee:2008fa}. For an
overview over the various available approaches and methods to nuclear
physics using chiral effective theories we refer to recent review
articles~\cite{Epelbaum:2008ga,Machleidt:2011zz}.

A lattice approach to nuclear physics was presented in a row of papers
by Borasay and
collaborators~\cite{Borasoy:2005yc,Borasoy:2006qn,Borasoy:2007vi,Epelbaum:2009zsa,Epelbaum:2009pd,Epelbaum:2010xt,Epelbaum:2011md}.
The authors of these publications take the approach of simulating the
chiral effective theory on a discrete space-time lattice. Starting
point for these simulations is the non-relativistic 
effective theory. As a consequence, Quantum Monte-Carlo methods based
on Fock states~\cite{Lee:2004qd} are used for the numerical
simulations. The lattice regularisation is utilised as the physical
cut-off in the theory, hence it cannot be removed. It naturally takes
values of $a\sim 2\ \mathrm{fm}$. Lattice artifacts are reduced
perturbatively by adding counter-terms to the simulated action. Such
large values of the lattice spacings allow for simulations with $L=8$
lattice points at most in each of the spatial directions. This is
important, as in the simulations a sign problem is apparent, which is
controllable for small $L$, but severe for large $L$.

Apart from the fact that QCD and the chiral effective field theory (ChET)
are rather different, also the approach taken in these publications
differs fundamentally from the methods applied in LQCD simulations.
In LQCD the path integral in Euclidean space-time is directly simulated
using appropriate Markov chain Monte Carlo methods.
Simulations are performed in a finite volume with discretised space-time,
with the inverse lattice spacing serving as a momentum cut-off.
Renormalisaton is then performed by properly removing the cut-off in
the {\em continuum limit} when the lattice spacing tends to zero ($a \to 0$).
At this point the fundamental difference between QCD and ChET becomes evident:
LQCD has a non-trivial continuum limit and hence the cut-off can be
completely removed.
In ChET  the cut-off cannot be removed -- in fact it has a physical relevance
representing the energy scale where the basic degrees of freedom (pions and
nucleons) become inappropriate for describing physics.
In a perturbative framework the necessity of a finite cut-off is implied
by the non-renormalisability of the theory.
In a non-perturbative formulation ChET belongs to the general class of
Yukawa-type theories with fermion-boson interactions.
(In a four-fermion theory, for instance, the boson in the Yukawa interaction
is generated dynamically.)
Yukawa-theories are expected to have only a trivial (non-interacting)
continuum limit, therefore the cut-off is necessary for maintaining a non-zero
interaction.
(For an introduction in lattice Yukawa models see Chapter 6 in 
\cite{montvay:1994ab}.)
The finite value of the cut-off also corresponds to the non-zero size of
pions and nucleons below which distance scale obviously some other description
is required (namely, with quarks and gluons in terms of QCD).

In the continuum limit of LQCD all symmetries of the continuum theory broken
by discretisation, including Lorentz symmetry, are restored and
lattice artifacts in all quantities are removed.
Performing the continuum limit requires an extrapolation to vanishing value of the
lattice spacing $a\to0$ and therefore the simulations have to take
place in ranges of $a$ which allow such an extrapolation in a 
controlled way.
It is not a priori clear how small values of the lattice spacing are required for
this procedure, and eventually one will only find it out empirically.

The so called Symanzik effective
theory~\cite{Symanzik:1981hc,Symanzik:1983dc,Symanzik:1983gh} allows
for a better understanding which lattice artifacts one has to expect.
For sufficiently small values of $a$ LQCD can be described by an
effective local action~\cite{Luscher:1996sc} 
\begin{equation}
  \label{eq:effS}
  S_\mathrm{eff} = S_0 + a S_1 + a^2 S_2 + \ldots\ ,
\end{equation}
where $S_0$ is the continuum QCD action and the additional terms $S_k$
for $k>0$ are to be interpreted as operator insertions in the
continuum theory
\[
S_k = \int d^4 x\ \mathcal{L}_k\, ,
\]
with $\mathcal{L}_k$ being combinations of local composite fields with
mass dimension $4+k$. The list of possible fields is constrained by
the symmetries. Similarly one can write for a local gauge invariant
composite field $\phi(x)$, where we neglect mixing under
renormalisation for simplicity, 
\begin{equation}
  \label{eq:effphi}
  \phi_\mathrm{eff}(x) = \phi_0(x) + a\phi_1(x) + a^2 \phi_2(x) + \ldots\ .
\end{equation}
The fields $\phi_k$ are again linear combinations of local fields with
appropriate symmetries and dimension. Without discussing all the
details -- which can be found in ref.~\cite{Luscher:1996sc} -- one
finds that a connected $n$-point function on the lattice to leading
order in the lattice spacing reads
\begin{equation}
  \label{eq:symanzik}
  \begin{split}
    G_n(x_1, & \ldots, x_n)  = \langle\phi_0(x_1)\cdots
    \phi_0(x_n)\rangle^\mathrm{cont} \\
    & - a\int d^4y\
    \langle\phi_0(x_0)\cdots\phi_0(x_n)\mathcal{L}_1(y)\rangle^\mathrm{cont} \\
    & + a \sum_{k=1}^n \langle \phi_0(x_1)\cdots
    \phi_1(x_k)\cdots\phi_0(x_n)\rangle^\mathrm{cont} \\
    & + \ldots\ .\\ 
  \end{split}
\end{equation}
From this expression it should be clear that first of all the value of
$a$ should be small enough to neglect all higher order terms in the
lattice spacing. In particular, on the right hand side of
eq.~(\ref{eq:symanzik}) there should be only combinations with
$a\langle\ldots\rangle \ll 1$. This is also the reason why LQCD
investigations with heavy quarks appear to be difficult since then 
$am_q$ can easily become of order $1$ and can lead to significant
lattice artifacts. Of course, all terms on the right hand side of
eq.~(\ref{eq:symanzik}) come with unknown coefficients, so also with
$am_q\approx 1$ large lattice artifacts are not automatically to be
expected.

The so called Symanzik expansion can also be used to systematically
remove lattice artifacts from the theory. This requires to add counter
terms to the lattice action with coefficients that need to be
determined non-perturbatively. Improving the action is, however, not
always enough. There are observables that require operator specific
improvement with coefficients that again need to be determined
non-perturbatively. 

Comparing the lattice QCD approach and the lattice approach to chiral
effective theories from 
refs.~\cite{Borasoy:2005yc,Borasoy:2006qn,Borasoy:2007vi,Epelbaum:2009zsa,Epelbaum:2009pd,Epelbaum:2010xt,Epelbaum:2011md}
rises an immediate question, namely how large are lattice artifacts in
the lattice simulations of the chiral effective theory and whether
they are will controlled. This question
is hard to answer since the continuum limit $a\to 0$ cannot be
performed due to the physical interpretation of the cut-off. Moreover,
a variation of the cut-off is for technical reasons difficult, but a
variation in a reasonable range is desirable.

This is why we take a different approach to lattice simulations of
these chiral effective theories, which we shall present in this
paper.
It consists of using the fully relativistic path integral formulation
of the chiral effective theory in Euclidean space-time, which can be
simulated by means of Markov chain Monte Carlo methods.
The physical cut-off of the theory is implemented by block-field methods.
The sizes of the blocks represent the hadron sizes.
These have to be kept fixed if the lattice spacing is changed, for
instance, keeping $R \cdot M$ fixed where $R$ denotes the block size
and $M$ is a physical mass, say the nucleon mass.
Separating the lattice cut-off from the physical cut-off allows to
change the lattice cut-off (i.e. the lattice spacing) by keeping the
physical content of the theory unchanged.
On the basis of the Symanzik effective theory it has to be expected that
by making the lattice spacing sufficiently small the lattice artifacts,
for instance, Lorentz-symmetry breaking can be suppressed.
One can also imagine to reach in the limit $a \to 0$ a non-trivial
continuum limit describing an interacting non-local quantum field theory.
Strictly speaking, the existence of such a continuum limit is not known
at present -- similarly to the existence of the continuum limit of many other
lattice quantum field theories.

A substantial simplification in fermionic lattice quantum field theories
is the so called \emph{quenched approximation} which corresponds to
omitting closed fermion loops.
Since both neutron and proton are heavy, we do not expect that
nucleon-anti-nucleon loops play a significant role and hence perform the
simulations in the quenched approximation.
Later on one can, of course, perform simulations in the full theory,
removing this approximation.

Another possible simplification is to work within a non-relativistic
approximation -- instead of in a relativistic quantum field theory as we
are doing here.
One big advantage of the non-relativistic formulation might be that
the rest masses of the nuclei do not appear in the theory and binding
energies are computed directly, whereas in our relativistic approach
binding energies are sub-percent effects as compared to the energy
levels we determine.
This is one point we shall address in a forthcoming publication.

The authors of refs.~\cite{DeSoto:2006jr,DeSoto:2006jt,deSoto:2011sy} 
follow a similar approach to the one presented here, but applied to a
Yukawa model with one scalar 
field. Previous investigations of Yukawa models in the Euclidean path
integral formulation have been performed in the context of upper and
lower limits of masses in the electroweak theory, see for instance
refs.~\cite{Farakos:1990ex,Lin:1993hp}.

In this paper we shall discuss our method and show evidence for its
applicability. Physical results are planned to be presented in
forthcoming publications. In the following section we discuss our
lattice action followed by details of the numerical methods we
apply. We shall close with presenting some simulation results and give a
conclusion and outlook.

\section{Lattice actions}
 
 Let us start by introducing our notations.
 The nucleon fields are described by a pair of Grassmann variables and are
 denoted by
\begin{equation}\label{eq2.1}
\psi_{\alpha x}\,, \hspace{2em}  \bar{\psi}_{\alpha x}  \hspace{2em}
(\alpha=1,2) 
\end{equation}
 where $\alpha$ is the isospin doublet index and $x$ denotes the lattice sites.
 The Dirac-index of the nucleon field is not displayed here and the isospin
 index of the nucleon field will also be omitted in most formulae.
 The real boson fields are
\begin{equation}\label{eq2.2}
\phi^{(\pi)}_{a x}\,, \hspace{2em} 
\phi^{(0)}_x\,,  \hspace{2em}  \phi^{(1)}_{a x}  \hspace{2em}
(a=1,2,3)
\end{equation}
 with the triplet isospin index $a$.
 Here $\phi^{(\pi)}$ stands for the pion and $\phi^{(0)}$ and $\phi^{(1)}$ are
 Hubbard-Stratanovich auxiliary boson fields with isospin zero and one,
 respectively.

 These latter are used to describe four-nucleon interactions which in the
 Euclidean lattice action have the following form:
\begin{equation}\label{eq2.3}
S_\mathrm{NA} = \sum_x \left\{
\phi^{(0)}_x\phi^{(0)}_x + \phi^{(1)}_{a x}\phi^{(1)}_{a x} +
C_0\, \phi^{(0)}_x\, (\bar{\psi}_x \psi_x) +
C_1\, \phi^{(1)}_{a x}\, (\bar{\psi}_x \tau_a \psi_x) \right\} \,,
\end{equation}
 where $\tau_a , (a=1,2,3)$ are Pauli-matrices for isospin and a summation
 over repeated indices $a$ is understood.
 The four-nucleon interactions are obtained after integrating over the
 auxiliary fields according to
\begin{equation}\label{eq2.4}
\int_{-\infty}^\infty d \phi\, 
\exp\{-\phi^2 - C\, \phi (\bar{\psi} \psi) \} =
\sqrt{\pi}\, \exp\{ \frac{C^2}{4}\, (\bar{\psi} \psi)^2 \} \,.
\end{equation}

 Our choice of the pion-nucleon interaction corresponds to the lattice
 discretisation of the continuum interaction
 $\partial_\mu\pi_a (\bar{\psi} \gamma_5\gamma_\mu \tau_a\psi)$ and is
 given as
\begin{equation}\label{eq2.5}
S_{\mathrm{N}\pi} = \sum_x \left\{
\frac{C_\pi}{2}\, \left( \phi^{(\pi)}_{a x+\hat{\mu}} - 
                       \phi^{(\pi)}_{a x-\hat{\mu}} \right)
(\bar{\psi}_x \gamma_5\gamma_\mu \tau_a \psi_x) \right\} \,,
\end{equation}
 where $\hat{\mu}$ denotes, as usual, the unit vector on the lattice in
 direction $\mu\, (\mu=1,2,3,4)$.
 (Similarly to isospin index $a$, over repeated direction index $\mu$
 a summation is implied.)
 The above expression corresponds to a particularly simple
 discretisation of the
 derivative of the pion field which can, of course, be chosen differently.

 The parameters of the lattice action are always dimensionless.
 The connection to the (eventually) dimensionful parameters in continuum
 formulations is established with the multiplication by an appropriate
 power of a mass parameter.

 Let us illustrate this on the example of the four-nucleon interactions
 in (\ref{eq2.3})-(\ref{eq2.4}).
 The (bare) nucleon field $\psi_x$ is related to the (bare) continuum
 nucleon field by $\psi_x = a^{3/2} \psi_x^\mathrm{cont}$ (and
 similarly for $\bar{\psi}_x$). The relation for the auxiliary scalar
 fields is given by $\phi_x = a^2 \phi_x^\mathrm{cont}$.
 This implies that the four-nucleon couplings $C_{0,1}$ are related to
 their continuum counterparts by $C_{0,1} = a^{-1} C_{0,1}^\mathrm{cont} =
 (m_\mathrm{N} C_{0,1}^\mathrm{cont})/(am_\mathrm{N})$.
 Here, $m_\mathrm{N} C_{0,1}^\mathrm{cont}$ is again dimensionless and
 independent of the lattice spacing $a$. In the following we always
 work in lattice units. The connection to the (eventually)
 dimensionful parameters in continuum formulations is established by
 multiplication with an appropriate power of a mass parameter and/or a
 power of the lattice spacing $a$. Note that in
 Ref.~\cite{Borasoy:2006qn} the coefficients appear quadratically in
 the action.
 
 In addition to the interaction terms we also need a kinetic term for the
 nucleon which we take to be the Wilson fermion lattice action \cite{Wilson:1974sk}:
 \begin{equation}\label{eq2.6}
   S_\mathrm{N} = \sum_x \left\{ 
     (\bar{\psi}_x \psi_x) - \kappa_\mathrm{N} \sum_{\mu=\pm 1}^{\mu=\pm 4}
     (\bar{\psi}_{x+\hat{\mu}} [1+\gamma_\mu] \psi_x) \right\} \,.
 \end{equation}
 Here $\kappa_\mathrm{N}$ is the hopping parameter defining the
 nucleon mass and the  convention $\gamma_{-\mu} = -\gamma_\mu$ is followed.
 For the pion field, besides the kinetic term, also a self-interaction term
 is allowed, hence we define the pion part of the action by
 \begin{equation}\label{eq2.7}
   S_{\pi} = \sum_x \left\{
     - 2 \kappa_\pi  \sum_{\mu=1}^{\mu=4} 
     \phi^{(\pi)}_{ax} \phi^{(\pi)}_{a x+\hat\mu} +
     \lambda [\phi^{(\pi)}_{ax} \phi^{(\pi)}_{ax} - 1]^2  \right\} \,,
 \end{equation}
 where $\kappa_\pi$ is the hopping parameter of the pion and $\lambda$ gives
 the strength of the self-interaction.
 The total lattice action is the sum of all the above terms, that is
\begin{equation}\label{eq2.8}
S = S_\mathrm{N} + S_\pi + S_\mathrm{NA} + S_{\mathrm{N}\pi} \,.
\end{equation}
%
   
\subsection{Block fields}

 The interaction among hadrons, like nucleons and pions, is non-local as a
 consequence of the extended hadron structure implied by Quantum
 Chromodynamics.  
 This non-locality can be approximately taken into account by introducing
 block fields in the lattice action.
 Block fields are sums of the the above local fields weighted by appropriate
 numerical factors.
 In order to decrease rotation symmetry breaking, our definition of the block
 fields tries to be as close as possible to an exponential decrease of the
 weight factors proportional to the Euclidean distance squared.
 Of course, periodic (or anti-periodic) boundary conditions have to be taken
 into account and therefore we define the squared distance between two points
 $x,y$ on the lattice as
\begin{equation}\label{eq2.9}
(x,y)^2 \equiv \sum_{\mu=1}^{\mu=4} |x_\mu,y_\mu|^2
\end{equation}
 where
\begin{equation}\label{eq2.10}
|x_\mu,y_\mu| \equiv \min 
\left(|x_\mu-y_\mu|,|x_\mu-y_\mu+L_\mu|,|x_\mu-y_\mu-L_\mu|\right) \,.
\end{equation}
 Here $L_\mu$ denotes the lattice extension in the direction $\mu$.
 With this definition a generic block field is defined by
\begin{equation}\label{eq2.11}
\Phi_x \equiv \sum_{y\,, (x,y)^2 \leq R^2} \phi_y\, 
\exp\left\{- S\; (x,y)^2\right\} \,.
\end{equation}
 The blocking parameters can depend on the type of fields, that is one can
 have different parameters for the nucleon
 $(R_\mathrm{N},S_\mathrm{N})$, for the pion 
 $(R_\pi,S_\pi)$ and for the  auxiliary fields $(R_0,S_0)$ and
 $(R_1,S_1)$, respectively. 
 The lattice action in terms of the block fields has exactly the same form
 as the above action in terms of the local fields (including the summation
 over the lattice sites).
 The only changes are:
\begin{equation}\label{eq2.12}
\psi_x \rightarrow \Psi_x \,,          \hspace{2em}
\phi_x^{(A)} \rightarrow \Phi_x^{(A)} \hspace{2em}
(A=\pi,0,1) \,.
\end{equation}

 Besides reflecting the non-locality of the interactions, the blocking also has
 an important role in defining the cut-off of the theory.
 The high momentum modes are decoupled from the interactions by the blocking.
 Only the modes below some momentum cut-off are interacting, the cut-off value
 being determined by the blocking parameters $(R,S)$.
 The cut-off introduced by the blocking can be considered as a physical effect.
 The lattice cut-off can be changed independently of it.
 This allows to move the lattice spacing to small values for reducing lattice
 artefacts.
 
\section{Numerical simulations}

\subsection{Lattice parameters}\label{sec3.1}

 In order to gain experience with the lattice formulation defined in the
 previous section we started numerical simulations on small lattices and also
 introduced some simplifications in the choice of lattice action parameters.
 In most cases we did not block the nucleon field, only the bosonic fields and
 fixed the blocking parameters as follows:
\begin{equation}\label{eq3.1}
R_\mathrm{N} = 0\,, \hspace{2em}
S_0 = S_1 = S_\pi = 2.0\,, \hspace{2em}
R_0 = R_1 = R_\pi = 1.5\,.
\end{equation}
 This choice of the radius sets the number of points in a block to be 33.
 
 The simplest choice of coupling parameters is to keep only $C_0$ and $C_1$
 defining the four-nucleon interactions.
 We also experimented with the additional introduction of the pion field and
 the coupling $C_\pi$ but here we report only on results in the pion-less theory.
 A detailed investigation of the effect of the pion coupling is left to future
 work.
 Following Ref.~\cite{Borasoy:2006qn} $C_0$ is taken to be real which corresponds
 to an attractive interaction whereas $C_1 = i\, |C_1|$ is purely imaginary
 describing a repulsive interaction.
 In most cases the magnitudes of $C_0$ and $C_1$ were equal, that is
\begin{equation}\label{eq3.2}
C_\pi = 0\,, \hspace{2em}
C_0 = |C_0|\,,\hspace{1em}  C_1 = i\,|C_1|\,,\hspace{1em} |C_0| = |C_1|\,.
\end{equation}
 (Note that according to eq.~(\ref{eq2.4}) the signs of $C_0$ and $C_1$ are
 irrelevant.)
 
 For this exploratory work we had access to a PC cluster with 12
 graphics processing units (GPUs) attached to it. Due to the memory
 limitations of the GPUs the maximally feasible spatial lattice size was 
 $32^3\,, (L=32)$.
 The temporal lattice extension is taken four times longer in order to allow
 for a precise determination of the energy values.
 In summary, we did simulations on three types of lattices:
\begin{equation}\label{eq3.3}
16^3 \cdot 64\,,  \hspace{2em}
24^3 \cdot 96\,,  \hspace{2em}
32^3 \cdot 128\,.
\end{equation}
This restriction to the currently feasible lattice volumes has
implications on the choice of parameters we can simulate: in order to
study the two nucleon system a large physical volume is required,
because otherwise the smallest available lattice momentum is too
large. On the other hand we would like to simulate with as small as
possible values of the lattice spacing $a$. 
To set the scale we fixed the equal mass of the proton and neutron to be
$M_\mathrm{N} \equiv M_{\rm proton} = M_{\rm neutron} = 939\, {\rm MeV/c^2}$.
This means that, for instance, a nucleon mass in lattice units $aM_\mathrm{N} = 1$
implies a lattice spacing $a \simeq 0.21\,{\rm fm}$ and in our cases spatial
extensions of $16\, a \simeq 3.4\,{\rm fm}$,  $24\, a \simeq 5.0\,{\rm fm}$
and $32\, a \simeq 6.7\,{\rm fm}$, respectively.
In order to reach larger volumes one can, of course, increase the nucleon mass
in lattice units but in this way lattice artefacts will also increase.
 
With these choices of volumes and with $aM_\mathrm{N} = 1$ we get for the
smallest non-zero momentum
$(2\pi)/(16 L a) \simeq 369\, {\rm MeV/c}$, 
$(2\pi)/(24 L a) \simeq 246\, {\rm MeV/c}$ and
$(2\pi)/(32 L a) \simeq 184\, {\rm MeV/c}$, respectively. For
extracting physics using the approach presented here both lattice
sizes and minimal non-zero momenta would require a factor, say, four
increase of the lattice extensions which will be comfortably possible
to reach with present day computer resources. 

\subsection{Numerical methods}

 Our present aim is to determine the energies (masses) of different nuclear
 systems.
 This can be achieved by investigating the large (Euclidean-) time behaviour
 of different sets of correlators.
 Since at present we restrict ourselves to quenched simulations, where the fermion
 determinant of the nucleon field is neglected, the creation of the boson field
 configurations is simple.
 In case of the auxiliary fields $\Phi_x^{(0)}$ and $\Phi_x^{(1)}$ one has to
 create Gaussian distributions.
 The pion field $\Phi_x^{(\pi)}$ can also be simply produced by some update
 algorithm as, for instance, Metropolis algorithm --  the only mild complication
 being to take into account the non-locality introduced by the blocking.
 
 The nucleon mass can be determined from the behaviour of the nucleon time-slice 
 correlator.
 The time-slice operators are defined as
\begin{equation}\label{eq3.4}
N_t \equiv 
\sum^{x_4=t}_{x_1\,,x_2\,,x_3} \Psi_{x_1\,,x_2\,,x_3\,,x_4} \,,
\hspace{2em}
\bar{N}_t \equiv 
\sum^{x_4=t}_{x_1\,,x_2\,,x_3} \bar{\Psi}_{x_1\,,x_2\,,x_3\,,x_4}
\end{equation}
 and the nucleon correlator is, with a Dirac-projection to the state propagating
 in positive direction,
\begin{equation}\label{eq3.5}
{\rm Tr_{Dirac}} 
\left[ (1+\gamma_4) \langle N_{t_1} \bar{N}_{t_2} \rangle \right] \,.
\end{equation}
 The expectation value of the fermion bilinear gives a fermion propagator
 which is the inverse of the fermion matrix in the fermionic part of the 
 action.
 The overwhelming part of computer resources in our quenched
 simulations is spent in the calculation of the
 fermion propagators by an iterative inverter of this sparse matrix.

 The computation of nucleon propagators has been done most of the time
 by applying a {\em mixed precision Conjugate Gradient} inverter, see
 the appendix.
 The crucial problem for the inverter is to deal with the very small 
 values of the nucleon propagators at large distances.
 The solution of this problem is to use {\em distance preconditioning}
 following Ref.~\cite{deDivitiis:2010ya}.
 Since the nucleon propagator behaves nearly exponentially for distances
 which we use for extracting the masses (in most cases up to a time
 distance half the time extension $L_4$ of the lattice), we choose the
 preconditioning function to be       
 \begin{equation}\label{eq3.6}
   \alpha(t) = \left\{
     \begin{array}{lll}
       \exp\{ -P\, t\}        & {\rm if} &  t \leq L_4/2 \,,  \\[1em]
       \exp\{ -P\, (L_4 - t)\}& {\rm if} &  t > L_4/2   \,.
     \end{array} \right .
 \end{equation}
 The parameter $P$ can be chosen typically by an amount $0.1-0.5$ smaller
 than the nucleon mass in lattice units $am_\mathrm{N}$.

 In order to obtain the masses of multi-nucleon (in the present paper
 two-nucleon) states with sufficient precision, one has to find the
 proper composite operators defining the correlators.
 Here we restrict ourselves to proton-neutron states.
 For local operators we take in the spin-0 and spin-1 channels,
 respectively,
 \begin{equation}\label{eq3.7}
   \Psi_{1x} C \gamma_5 \Psi_{2x} \,, \hspace{2em}
   \Psi_{1x} C \gamma_k \Psi_{2x} \,, \hspace{2em} (k=1,2,3) \,,
 \end{equation}
 where $C$ denotes the charge conjugation Dirac matrix.

 Especially for scattering states it is important to also take
 extended (smeared) operators where the proton and neutron are at    
 different points. 
 In case of {\em Gaussian smearing} one can use the smearing function
\begin{equation}\label{eq3.8}
\exp\left\{ -\sigma_1 |x_1,y_1|^2 - \sigma_2 |x_2,y_2|^2 
       -\sigma_3 |x_3,y_3|^2 \right\}
\end{equation}
 with the notation introduced in (\ref{eq2.10}).
 For a spherical state in the spin-0 channel one can take
 $\sigma \equiv \sigma_1 = \sigma_2 = \sigma_3$.
 For spin-1, ellipsoidal states with e.g. 
 $\sigma_1 \ne \sigma_2 = \sigma_3$ are useful.
 In this latter case we also tried linear smearing corresponding to
 $\sigma_2,\sigma_3 = \infty$.
 In order to save computer time one can cut the summation over
 sites off at distances where the smearing function in (\ref{eq3.8})
 is smaller than, say, $10^{-2}$.
 In case of spherical smearing this corresponds to a cut-off radius of
\begin{equation}\label{eq3.9}
\rho = \left\{ \frac{\log(100)}{\sigma} \right\}^{1/2} \,.
\end{equation}

 The simplest way to determine the masses is to fit some of the
 correlators by an exponential function in time intervals for distant
 time-slices.
 In case of small enough statistical errors one can also obtain good
 fits with a sum of two (or more) exponentials.
 The best results can be achieved, however, by taking a set of some
 operators in a given channel and calculate the {\em correlator matrix}
 among them.
 For determining the energies of two-nucleon (actually proton-neutron)
 states we typically start from a $4 \times 4$ correlator matrix.
 The four states are chosen from local, spherically smeared and elliptically
 smeared states with different Dirac-matrices.

 The correlator matrix can be approximated by the sum of contributions
 of eigenstates of the Hamiltonian (i.~e. of the transfer matrix).
 In general, a real symmetric $D \times D$ correlator
 matrix $C(t_2,t_1)$ between time-slices $t_1$ and $t_2 > t_1$ 
 is defined by the matrix elements of $D$ operators
 ${\cal O}_a,{\cal O}_b,\ldots,{\cal O}_d$.
 If the energy eigenstates are $|n\rangle,\; n=1,2,\ldots,M$ then in a
 shorthand notation
\begin{equation}\label{eq3.10}
C(t_2,t_1) = \left(
\begin{array}{cccc}
C(t_2,t_1)_{aa} & C(t_2,t_1)_{ab} & \ldots & C(t_2,t_1)_{ad} \\
C(t_2,t_1)_{ab} & C(t_2,t_1)_{bb} & \ldots & C(t_2,t_1)_{bd} \\
\vdots          & \vdots          & \ldots & \vdots          \\
C(t_2,t_1)_{ad} & C(t_2,t_1)_{bd} & \ldots & C(t_2,t_1)_{dd}
\end{array}
\right)
\end{equation}
 where the matrix elements can be written as, for instance,
\begin{equation}\label{eq3.11}
C(t_2,t_1)_{ab} =
(a|1)_{t_2}(b|1)_{t_1} + (a|2)_{t_2}(b|2)_{t_1} + \ldots + (a|M)_{t_2}(b|M)_{t_1}
\end{equation}
 with
\begin{equation}\label{eq3.12}
 (c|k)_t \equiv
\langle 0 | {\cal O}_c(t) | k \rangle = \langle k | {\cal O}_c(t) | 0 \rangle \ ,
\end{equation}
 for $c=a,b,\ldots,d$ and $k=1,2,\ldots,M$.

 Assuming that we consider bosonic (fermionic) operators, we have periodic
 (anti-periodic) time dependence with the time extension of the lattice $L_4$.
 This implies
\begin{equation}\label{eq3.13}
(a|k)_{t_2} (b|k)_{t_1}  = (a|k)\,(b|k) 
\left\{ \exp[-t E_k] \pm \exp[-(L_4-t) E_k] \right\}
\ .
\end{equation}
 where the positive and negative sign stands for periodicity and
 anti-periodicity, respectively.
 Here $t \equiv t_2-t_1$, $E_k$ is the energy (e.g. mass) corresponding 
 to the state
 $|k\rangle$ and \\
 $(a|k) \equiv (a|k)_0,\; (b|k) \equiv (b|k)_0$.
 Fitting the correlator matrix by the expression given by 
 (\ref{eq3.10}) - (\ref{eq3.13}) one can obtain the energies we are 
 looking for \cite{Baron:2010th}.
 The statistical errors of the results can also be obtained by methods
 similar to those described in Section 5 of this reference.

 Since in the present case the relevant (multi-) nucleon correlators
 can be determined to a very good precision, one can perform least-square
 fits by minimising the correlated chi-squared.
 In order to obtain a good starting point for the minimisation, one
 can first minimise the {\em uncorrelated chi-squared} defined by
\begin{equation}\label{eq3.14}
\chi_n^2 = \sum_{i=1}^{N_C} \left(\frac{f_i(p) -
    \overline{X}_i}{\delta X_i}\right)^2 
\end{equation}
 where the index $i$ runs over the independent matrix elements to be fitted,
 $\overline{X}_i$ and $\delta X_i$ are the mean value and error of the
 matrix element $i$, respectively, and $f_i(p)$ is the fitting function of $N_P$ 
 parameters $(p_1,p_2,\ldots,p_{N_P})$ defined by (\ref{eq3.11})-(\ref{eq3.13}).
 The best fit obtained in this way can be taken as a starting point
 to minimise the {\em correlated chi-squared} 
\begin{equation}\label{eq3.15}
\chi_c^2 = \sum_{i,j=1}^{N_C} \left(f_i(p) - \overline{X}_i\right) M_{ij}
                              \left(f_j(p) - \overline{X}_j\right) \ ,
\end{equation}
 where $M_{ij} = N C_{ij}^{-1}$, with the number $N$ of input data and
 the correlator matrix
\begin{equation}\label{eq3.16}
C_{ij} = \frac{1}{N-1}\sum_{n=1}^{N} \left(X_{i,n} - \overline{X}_i\right)
                              \left(X_{j,n} - \overline{X}_j\right) \ .
\end{equation}

 In general, the correlator matrix in (\ref{eq3.16}) can be determined
 with sufficient precision for obtaining its inverse and its eigenvectors.
 In some cases, in particular if the dimension of the correlator matrix
 $N_C$ is large, {\em smoothing} of the smallest eigenvalues
 \cite{Michael:1993yj,Michael:1994sz} can be helpful but does not
 substantially influence the results.
 The advantage of properly obtaining the minimum of $\chi_c^2$ is that
 one can select ``good fits'' by the value of $\chi_c^2$ per number of
 degrees of freedom $(N_C-N_P)$.
 The mean value and error of a quantity is defined by considering the
 distribution of its values in good fits.
 The quoted value is then the position of the median of the distribution
 of these selected values.
 The error defines a (symmetric) interval around the median such that 68\% of
 the distribution is contained in it.

\begin{figure}[t]
  \begin{center}
    \subfigure[]{\includegraphics[width=0.45\linewidth]%
      {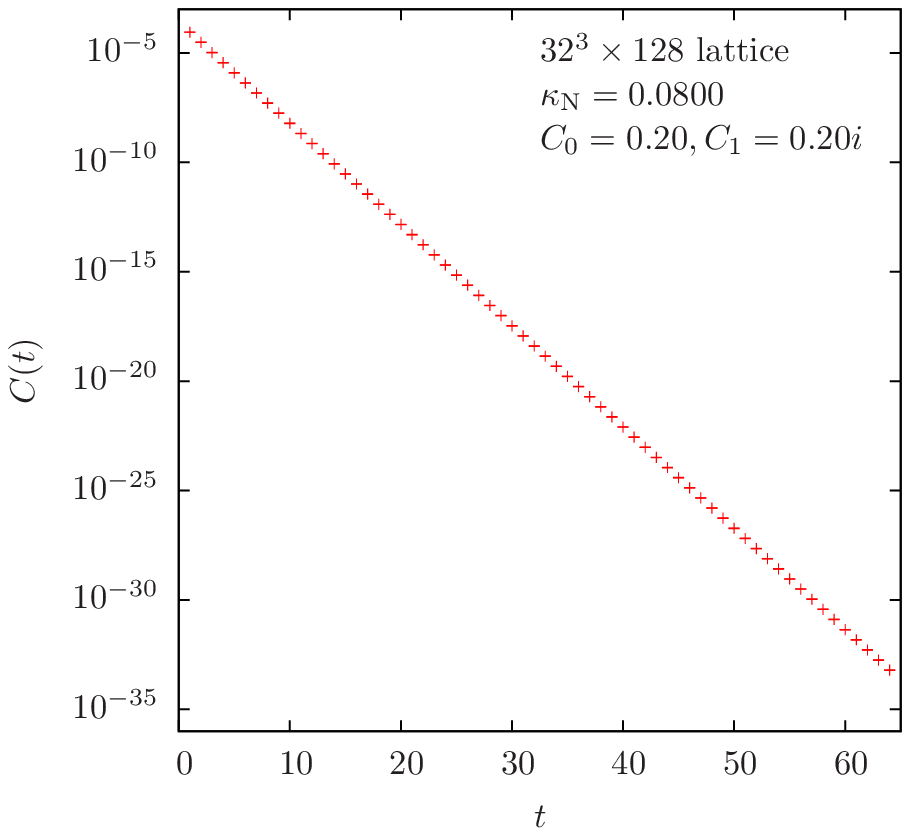}}
    \qquad
    \subfigure[]{\includegraphics[width=0.45\linewidth]%
      {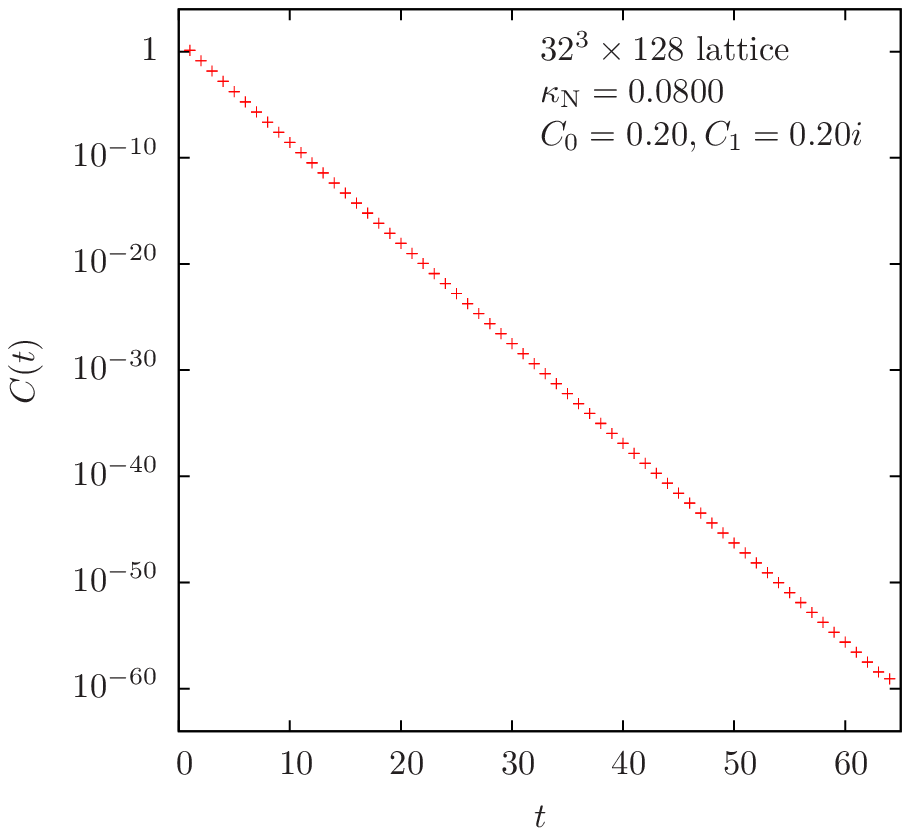}}
  \end{center}    
  \caption{\label{fig01}\em
    The nucleon correlator (a) and two-nucleon correlator (b)
    on a logarithmic scale as a function of time-slice distance $t$ in
    lattice units on a $32^3
    \cdot 128$ lattice at $\kappa_\mathrm{N}=0.08,\; C_0=-iC_1=0.2$.}
\end{figure}
\subsection{Numerical results}

The physical quantities we are interested in are for instance
the nucleon-nucleon scattering length and binding energies of multi
nucleon states. In a lattice simulation, the determination of these
quantities requires a study of the (finite) volume dependence of one
and two (and multiple) particle energies. In this methodical paper we
hence try to understand how precisely the corresponding quantities,
i.e. the nucleon and two-nucleon masses, can be determined. 

\begin{figure}[t]
  \vspace*{0.01\vsize}
  \begin{center}
    \begin{minipage}[c]{1.0\linewidth}
      \hspace{0.10\hsize}
      \includegraphics[width=0.80\hsize]
      {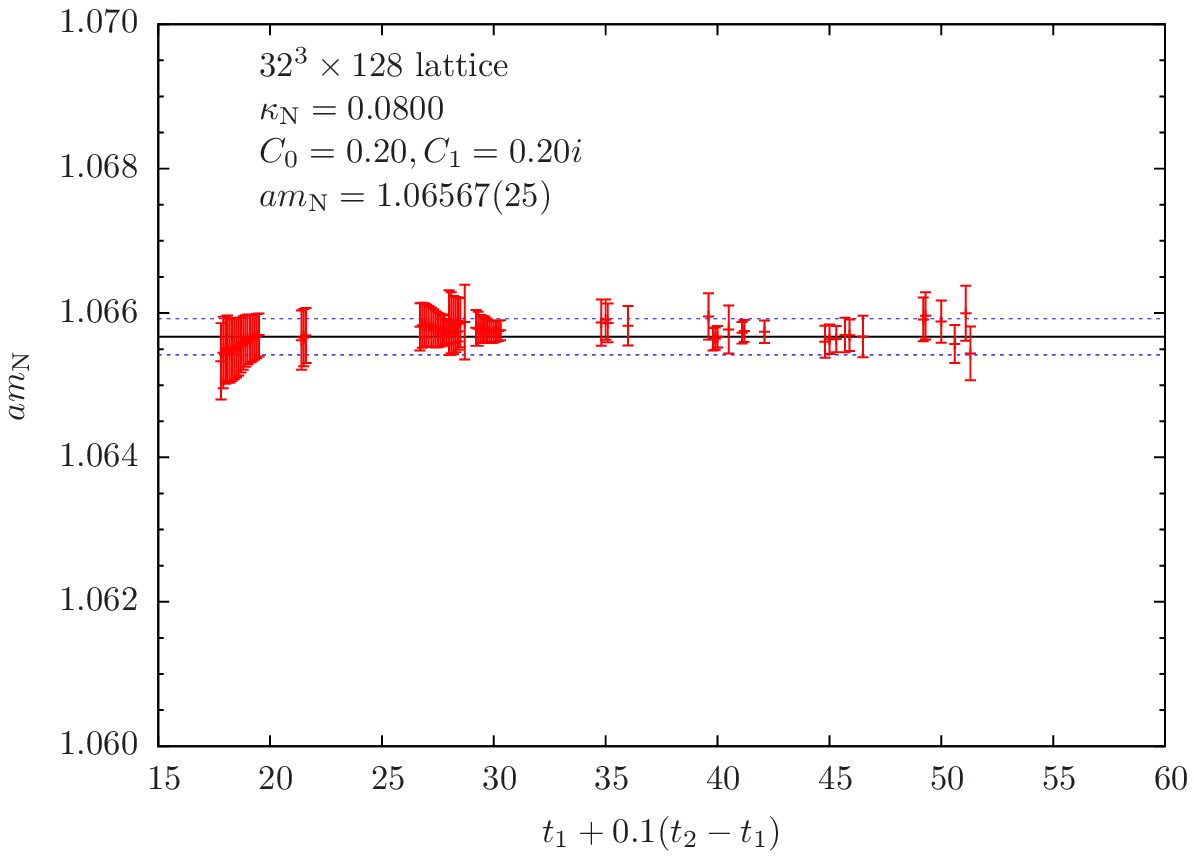}
    \end{minipage}
  \end{center}
  \caption{\label{fig02}\em
    Results of nucleon mass $am_\mathrm{N}$ fits on different fitting
    intervals for a $32^3 \cdot 128$ lattice at $\kappa_N=0.08,\;
    C_0=-iC_1=0.2$. The fit interval $[t_1,t_2]$ is specified on the
    $x$-axis by $t_1+0.1 (t_2-t_1)$.
    The horizontal lines indicate the final result and error obtained from
    the distribution of the fit results.}
\end{figure}

\begin{figure}[t]
  \vspace*{0.01\vsize}
  \begin{center}
    \begin{minipage}[c]{1.0\linewidth}
      \hspace{0.10\hsize}
      \includegraphics[width=0.80\hsize]
      {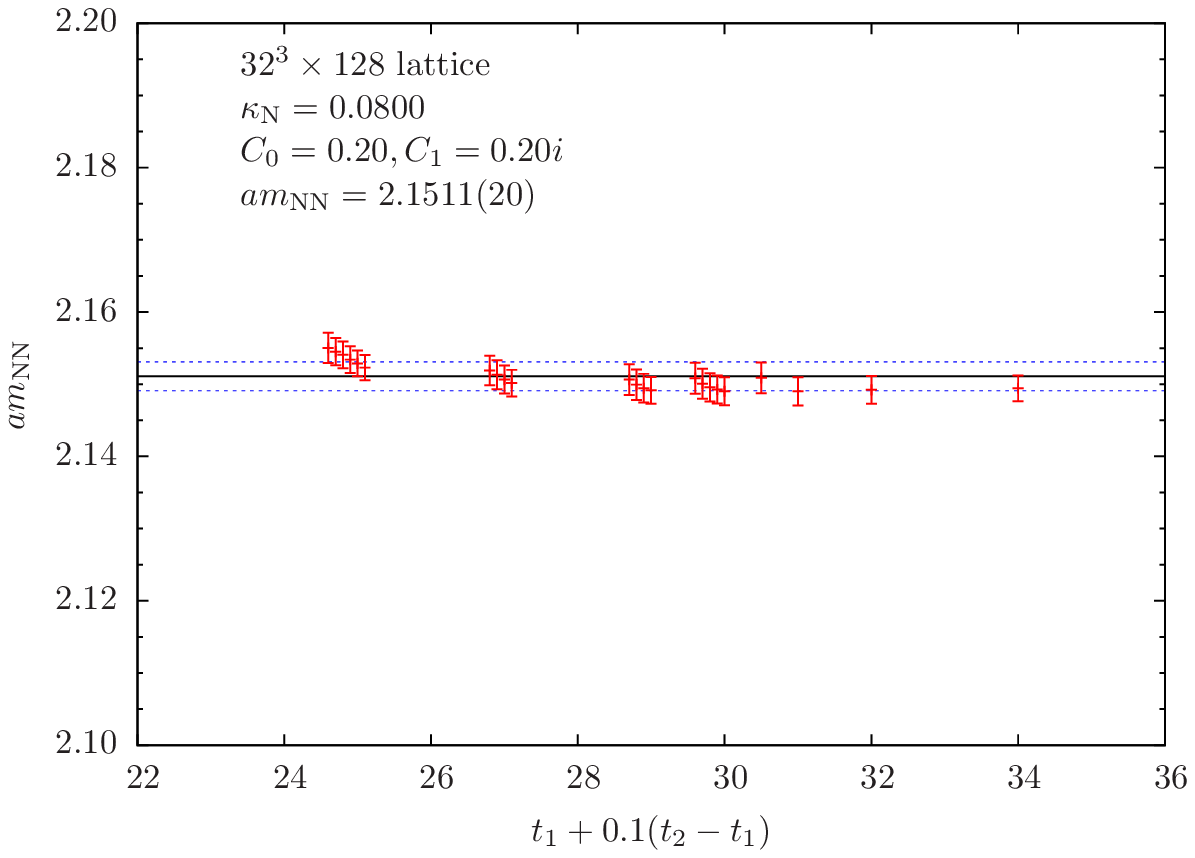}
    \end{minipage}
  \end{center}
  \caption{\label{fig03}\em
    The same as Fig.\protect{\ref{fig02}} for two-nucleon
    masses $am_\mathrm{NN}$.}
\end{figure}

In order to do so, we performed several simulations and determined the
masses as described in the previous sub-section. A typical example is
a run on a $32^3 \cdot 128$ lattice at $\kappa_\mathrm{N}=0.08,\;
C_0=0.2,\; C_1=0.2i$. It turned out that the masses can be very
precisely obtained even from a modest statistical sample of 120
configurations: see figures \ref{fig01}-\ref{fig03}.

In figure \ref{fig01} we plot the actual nucleon and two nucleon
correlators as functions of the time $t$ in lattice units on a
logarithmic scale. The decay is nearly exponential in the whole time
range.
Fitting the correlators in different time-slice distance intervals
$[t_1,t_2]$ by minimising the correlated chi-squared one finds that 
for the nucleon mass $am_\mathrm{N}$ we observe a plateau from $t_1=18$ on, 
almost independent of the value of $t_2$ (see fig. \ref{fig02}). 
For the two nucleon mass $am_\mathrm{NN}$ the plateau sets in somewhat
later (see fig. \ref{fig03}).
Both quantities have in common that the statistical errors of
the single points are in the sub-percent region. 
The mass value and its error as determined from the distribution of the
fit results -- as described earlier -- are indicated in both plots by 
the horizontal lines.

 The required precision of two-nucleon energies can be exemplified
 by L{\"u}scher's formula for extracting scattering lengths from
 finite size effects of the two-particle energies~
 \cite{Luscher:1986pf}. 
 To leading order one can express the scattering length $a_0$ as
\begin{equation}\label{eq3.17}
a_0 m_\mathrm{N} = \left( 2-\frac{m_\mathrm{NN}}{m_\mathrm{N}} \right) \cdot
\frac{(m_\mathrm{N} La)^3}{4\pi} \ .
\end{equation}
 The masses from Figures \ref{fig02}-\ref{fig03}, namely
 $am_\mathrm{N} = 1.06567(25)$ and $am_\mathrm{NN} = 2.1511(20)$ give for the right hand
 side a value $-59(7)$.
 The physical value of the right hand side is $a_0 m_\mathrm{N} = 113.1$.
 This corresponds to the value of the scattering length in the $^1S_0$
 channel $a_0 = +23.76\,{\rm fm}$ and $m_\mathrm{N} = 939\,{\rm MeV}$.
 (Here we use the sign convention for the scattering length of
 Ref.~\cite{Luscher:1986pf}.\footnote{We thank the referee of our paper
 for drawing our attention to the different sign conventions of the
 scattering length in the literature which we overlooked.})
 Our value has the right order of magnitude but an opposite sign which is due
 to the strong repulsion implied by the imaginary value $C_1=0.2i$.
 Obviously, tuning to the physical value would require either a much smaller
 imaginary value or even a real value of $C_1$.
 Besides of this, one cannot assume that the asymptotic formula works well
 already on a volume of extension $\simeq 7\,{\rm fm}$.
 In fact, on lattices $16^3 \cdot 64$ and $24^3 \cdot 96$ in the same
 point we obtained for the right hand side of (\ref{eq3.17}) values of
 about 5 and 40, respectively.
 This shows that, in any case, for the determination of scattering lengths
 simulations on larger volumes are required.

\section{Outlook}

 In this paper we have defined a theory of nucleon and pion fields in
 the Euclidean path integral formulation on the lattice.
 The inherent physical cut-off of this theory has been implemented
 by formulating the lattice action in terms of blocked fields.
 This physical cut-off is given by the block size.
 In this way the lattice cut-off is separated from the physical
 cut-off and can be changed in order to determine the size of
 lattice artefacts.

 The positive outcome of the first studies we have performed is that
 the correlators of single- and two-nucleon systems can be
 determined very precisely in order to obtain the energies
 to a very good precision.
 In fact a few thousands of configurations are sufficient for
 a precision better than one per mill. Using L{\"u}scher's formula we
 could determine values for the nucleon-nucleon scattering length. The
 corresponding results are positive in the sense that with larger
 volumes it seems to be realistic to tune its value to the physical
 one. 

 An important step towards obtaining physical results is hence to
 increase the physical lattice sizes.
 As discussed in Section \ref{sec3.1}, a spatial lattice extension
 of about $L=100$ would correspond for a nucleon mass in lattice units
 $aM_\mathrm{N} \simeq 1$ to a lattice size of about $La \simeq
 20\,{\rm fm}$ and 
 a minimal lattice momentum $2\pi/(100 La) \simeq 60\,{\rm MeV/c}$.

 It will be also important to introduce the pion field besides the
 auxiliary fields describing four-nucleon couplings.

 In order to complete this sort of lattice studies an important
 final step is to investigate the dependence of the results on the
 lattice spacing.
 For this the physical parameters as, for instance, the block size
 parameters $Rm_\mathrm{N}$, $Sm_\mathrm{N}$ and the lattice volume $Lam_\mathrm{N}$ have to be kept
 fixed.
 The couplings in the lattice action ($C_{0,1}$ etc.) have to be
 tuned for each value of the lattice spacing in such a way that some
 well chosen physical quantities (as for instance some nucleon phase
 shifts) take their physical values.
 Of course, if the lattice spacing gets smaller the required number
 of lattice points have to be increased correspondingly and this
 implies an increase in the required computational power. 

 A possible source of difficulties in the quenched approximation,
 as observed in Yukawa models by the authors of
 \cite{DeSoto:2006jr,DeSoto:2006jt,deSoto:2011sy}, is the appearance
 of {\em exceptional configurations} with extremely small eigenvalues
 of the fermion lattice action.
 These configurations make the determination of correlators and
 therefore masses practically impossible. 
 In our case we found exceptional configurations for bare couplings
 in the range $|C_0|,\, |C_1|,\, |C_\pi| > 0.3$.
 Since this problem does not appear in numerical simulations in
 Yukawa models with dynamical fermions \cite{Farakos:1990ex,Lin:1993hp},
 we expect that it does also disappear in our nuclear Yukawa models
 if dynamical nucleons are included in the simulation update.
 For real values of the couplings $C_0,\, C_1,\, C_\pi$ the fermion
 determinant is real (non-negative) therefore the known Hybrid
 Monte Carlo methods \cite{Duane:1987de} can be applied in a
 straightforward manner.
 For non-real (e.g. imaginary) couplings the determinant becomes
 complex and the numerical simulation turns non-trivial, if not
 impossible.

\vspace*{3em}\noindent
{\large\bf Acknowledgement}

\noindent
 We thank Ulf G. Mei{\ss}ner for introducing us in the literature of
 nuclear physics on the lattice. We thank Hans-Werner Hammer, Dean 
 Lee and Ulf G. Mei{\ss}ner for helpful and interesting
 discussions. We are grateful to Hans-Werner Hammer for useful
 comments on the manuscript. We thank JSC at FZ-J{\"u}lich for
 providing computing time on JUROPA.

 \begin{appendix}
   \section{Implementation Details}

As mentioned in the introduction, we have used graphics processing
units (GPUs) in order to perform the numerical inversions of the Dirac
operator. We have 12 NVIDIA Tesla C1060 GPUs available with four Gb of
memory each. We have used NVIDIAs CUDA environment to implement the
Dirac operator deduced from eqs.~(\ref{eq2.3}), (\ref{eq2.5}) and
(\ref{eq2.6}) for GPUs, which is very 
similar to available implementations for lattice QCD, see for instance
ref.~\cite{Clark:2009wm}. 

We employ a mixed precision solver using both, the CPU and the GPU. On
the GPU we have implemented a conjugate gradient (CG) solver inverting
the squared hermitian Dirac operator (since $C_1$ is purely imaginary)
\[
Q^\dagger Q = \gamma_5 D(C_0, -C_1, \kappa_N) \gamma_5 D(C_0, C_1,
\kappa_N)\ .
\]
The desired result is then obtained by multiplying with
$Q^\dagger$. The CG solver on the GPU is implemented solely in single
precision (32 Bit). The CG solver is called from an outer solver,
which is run on the CPU in double precision (64 Bit). We use iterative
refinement as the outer solver in order to solve
\[
D \eta = \phi
\]
for $\eta$, given some source spinor field $\phi$. The algorithm is
summarised in algorithm~\ref{alg:itref}. Depending on the parameters
$\kappa_N, C_0$ and $C_1$ one has to tune the precision for which to
solve on the GPU. The usage of distance preconditioning also had
significant influence on this tuning: the closer the preconditioning
mass to the measured mass the less stable the prescribed mixed
precision solver turned out to behave. Most probably due to
accumulation of round off errors we had to reduce the number of
inner iterations further and further with preconditioning mass
approaching the measured mass. 

\begin{algorithm}[t!]
  \caption{Iterative Refinement}
  \label{alg:itref}
  \begin{algorithmic}[1]
    \Require $\phi, \eta, \epsilon_o>0, \epsilon_i>0$
    \State $k=0$
    \State $r_k = \phi - D \eta$
    \While{$\|\phi-Dx_k\| > \epsilon_o$}
    \State solve $D p_{k+1} = r_k$ for $p_{k+1}$ on GPU to relative precision $\epsilon_i$
    \State $x_{k+1} = x_k + p_{k+1}$
    \State $r_{k+1} = \phi - D x_{k+1}$
    \State $k = k+1$
    \EndWhile
    \State \Return $\phi$
  \end{algorithmic}
\end{algorithm}


 \end{appendix}

\newpage
\bibliographystyle{h-physrev5}
\bibliography{nucleons}

\end{document}